\pdfoutput=1
\documentclass[11pt,twoside,a4paper,cmspaper,final,collab]{cms-tdr}
\def\svnVersion{3e4727f}\def\svnDate{2024/03/25}\def\cmsCernNoTag{CERN-EP-2023-256}\def\cmsCernDate{\today}\def\cmsMessage{Published in the Journal of High Energy Physics as \href{http://dx.doi.org/10.1007/JHEP03(2024)105}{\doi{10.1007/JHEP03(2024)105}.}}
\begin{document}\cmsNoteHeader{EXO-21-013}

\cmsNoteHeader{EXO-21-013}

\newcommand{\ctau}{\ensuremath{c\tau_{0}}\xspace}
\newcommand{\lept}{\Pell}
\newcommand{\hnl}{\PN}
\newcommand{\mhnl}{\ensuremath{m_{\hnl}}\xspace}
\newcommand{\vl}{\ensuremath{V_{\lept\PN}}\xspace}
\newcommand{\ve}{\ensuremath{V_{\Pe\PN}}\xspace}
\newcommand{\vmu}{\ensuremath{V_{\PGm\PN}}\xspace}
\newcommand{\vtau}{\ensuremath{V_{\PGt\PN}}\xspace}

\newcommand{\fl}{\ensuremath{\text{f}_{\lept}}\xspace}
\newcommand{\fe}{\ensuremath{\text{f}_{\Pe}}\xspace}
\newcommand{\fmu}{\ensuremath{\text{f}_{\PGm}}\xspace}
\newcommand{\ftau}{\ensuremath{\text{f}_{\PGt}}\xspace}

\newcommand{\leptI}{\ensuremath{\lept_{1}}\xspace}
\newcommand{\leptII}{\ensuremath{\lept_{2}}\xspace}
\newcommand{\eiso}{\ensuremath{I_{\text{rel}}(\Pe)}\xspace}
\newcommand{\muiso}{\ensuremath{I_{\text{rel}}(\PGm)}\xspace}
\newcommand{\mll}{\ensuremath{m_{\lept\lept}}\xspace}
\newcommand{\jn}{\ensuremath{\text{j}^{\star}}\xspace}
\newcommand{\mllj}{\ensuremath{m_{\lept\lept\jn}}\xspace}
\newcommand{\dxy}{\ensuremath{d_{\text{xy}}}\xspace}
\newcommand{\dxyerr}{\ensuremath{\sigma_{\text{xy}}}\xspace}
\newcommand{\dxysig}{\ensuremath{d_{\text{xy}}^{\text{sig}}}\xspace}
\newcommand{\Lxy}{\ensuremath{L_\text{xy}}\xspace}

\newcommand{\wjets}{\ensuremath{\PW\text{+jets}}\xspace}
\newcommand{\zjets}{\ensuremath{\PZ/\PGg^{*}\text{+jets}}\xspace}
\newcommand{\vgjets}{\ensuremath{\PV\PGg^{*}\text{+jets}}\xspace}

\newcommand{\pj}{\ensuremath{P(\jn)}\xspace}
\newcommand{\pql}{\ensuremath{P_{\PQq,\lept}(\jn)}\xspace}
\newcommand{\pq}{\ensuremath{P_{\PQq}(\jn)}\xspace}
\newcommand{\pe}{\ensuremath{P_{\Pe}(\jn)}\xspace}
\newcommand{\pmu}{\ensuremath{P_{\PGm}(\jn)}\xspace}

\DeclareRobustCommand{\PGmmp}{{\HepParticle{\PGm}{}{\mp}}\Xspace} 
\DeclareRobustCommand{\leptpm}{{\HepParticle{\Pell}{}{\pm}}\Xspace} 
\DeclareRobustCommand{\leptmp}{{\HepParticle{\Pell}{}{\mp}}\Xspace} 
\newcommand{\keVns}{\ensuremath{\text{ke\hspace{-.08em}V}}\xspace}

\title{Search for long-lived heavy neutral leptons with lepton flavour conserving or violating decays to a jet and a charged lepton}

\date{\today}

\abstract{
	
A search for long-lived heavy neutral leptons (HNLs) is presented, which considers the hadronic final state and coupling scenarios involving all three lepton generations in the 2--20\GeV HNL mass range for the first time. Events comprising two leptons (electrons or muons) and jets are analyzed in a data sample of proton-proton collisions, recorded with the CMS experiment at the CERN LHC at a centre-of-mass energy of 13\TeV, corresponding to an integrated luminosity of 138\fbinv. A novel jet tagger, based on a deep neural network, has been developed to identify jets from an HNL decay using various features of the jet and its constituent particles. The network output can be used as a powerful discriminating tool to probe a broad range of HNL lifetimes and masses. Contributions from background processes are determined from data. No excess of events in data over the expected background is observed. Upper limits on the HNL production cross section are derived as functions of the HNL mass and the three coupling strengths \vl to each lepton generation \lept and presented as exclusion limits in the coupling-mass plane, as lower limits on the HNL lifetime, and on the HNL mass. In this search, the most stringent limit on the coupling strength is obtained for pure muon coupling scenarios; values of $\abs{\vmu}^{2}>5\,(4)\times10^{-7}$ are excluded for Dirac (Majorana) HNLs with a mass of 10\GeV at a confidence level of 95\% that correspond to proper decay lengths of 17\,(10)\unit{mm}.
}

\hypersetup{
pdfauthor={CMS Collaboration},
pdftitle={Search for long-lived heavy neutral leptons with lepton flavour conserving or violating decays to a jet and a charged lepton},
pdfsubject={CMS},
pdfkeywords={CMS, long-lived particles, machine learning}}

\maketitle

\section{Introduction}

The observation of neutrino oscillations~\cite{SuperKamiokande:1998kpq,SNO:2002tuh,Esteban:2018azc}, a mechanism by which neutrinos can change their flavour, implies that at least two of the standard model (SM) neutrinos must have a nonzero mass and that neutral lepton flavour number is not a conserved quantity. Upper limits on the sum of all neutrino masses have been derived from cosmological considerations, whereas the study of particle decays via charged currents (\eg $\beta$-decay of tritium~\cite{KATRIN:2021uub}) has allowed upper limits to be set on the mass per lepton generation~\cite{Zyla:2020zbs}. The smallness of the neutrino masses compared to those of the other SM particles and to the vacuum expectation value of the Higgs potential suggests that there might exist another mechanism, beyond the SM, to explain the fundamental nature of neutrinos.

In this context, a minimal extension to the SM is given by a ``seesaw'' mechanism~\cite{seesaw1,seesaw2,seesaw3,seesaw4,seesaw5,seesaw6}, which postulates additional neutral leptons that are singlets with respect to the SM gauge groups and hence can possess an arbitrary mass. Through coupling to the Higgs field, these new particles mix with the SM neutrinos after electroweak symmetry breaking. The resulting mass eigenstates correspond to the three SM neutrinos with small masses and to three nearly sterile heavy neutral leptons (HNLs) that couple to the \PW and \PZ bosons in a similar way as the SM neutrinos. However, their coupling strengths are suppressed because of small mixing angles. An overview of the phenomenology of HNLs at the \GeV mass scale can be found in Ref.~\cite{Bondarenko:2018ptm}.

A common model that incorporates HNLs via the seesaw mechanism is the neutrino minimal standard model~\cite{Asaka:2005pn,Asaka:2005an}. Models incorporating HNLs are further motivated since, beside neutrino oscillations, they can explain additional phenomena that are not described by the SM: the baryonic asymmetry of the universe and the nature of dark matter~\cite{Boyarsky:2009ix,Drewes:2013gca}. Phenomenologically, an HNL can be either a Dirac or Majorana particle, \hnl, that is characterized by its mass \mhnl and coupling strengths \ve, \vmu, and \vtau to electrons, muons, and tau leptons, respectively. The HNL can be long-lived for sufficiently low coupling strengths since its partial three-body decay widths are proportional to $\mhnl^{5}\abs{\vl}^2$, where $\lept\in\{\Pe,\PGm,\PGt\}$~\cite{Bondarenko:2018ptm}. In general, lepton flavour number is not a conserved quantity in HNL interactions. It is only conserved in the special case where only one \vl coupling is nonzero.

Searches for HNLs are pursued at various experiments covering a wide range of masses from a few {\keVns} up to several {\TeVns}~\cite{Deppisch:2015qwa,Beacham:2019nyx}. At the CERN LHC, searches for HNLs are performed employing their production in the decay of mesons or through on-shell \PW or \PZ bosons, where the most stringent limits on \ve and \vmu can be derived starting at the \GeV mass scale. Corresponding analyses have been conducted by the ATLAS, CMS, and LHCb Collaborations~\cite{ATLAS:2019isd,ATLAS:2019kpx,ATLAS:2022atq,ATLAS:2023tkz,ATLAS:2023cjo,CMS:2018agk,CMS:2018iaf,CMS:2018jxx,CMS:2018iye,CMS:2022fut,CMS:2022hvh,CMS:2022chw,LHCb:2020wxx,LHCb:2020akw}, which mainly target signatures arising from the production of HNLs and their prompt decay to only one lepton generation (\Pe or \PGm), whereas long-lived HNLs and the possibility of cross-generation lepton couplings are only recently being explored. Various of these experimental results have been statistically combined using the \textsc{gambit} tool~\cite{Chrzaszcz:2019inj}.

\begin{figure*}[htb!]
\centering
\includegraphics[width=0.45\textwidth]{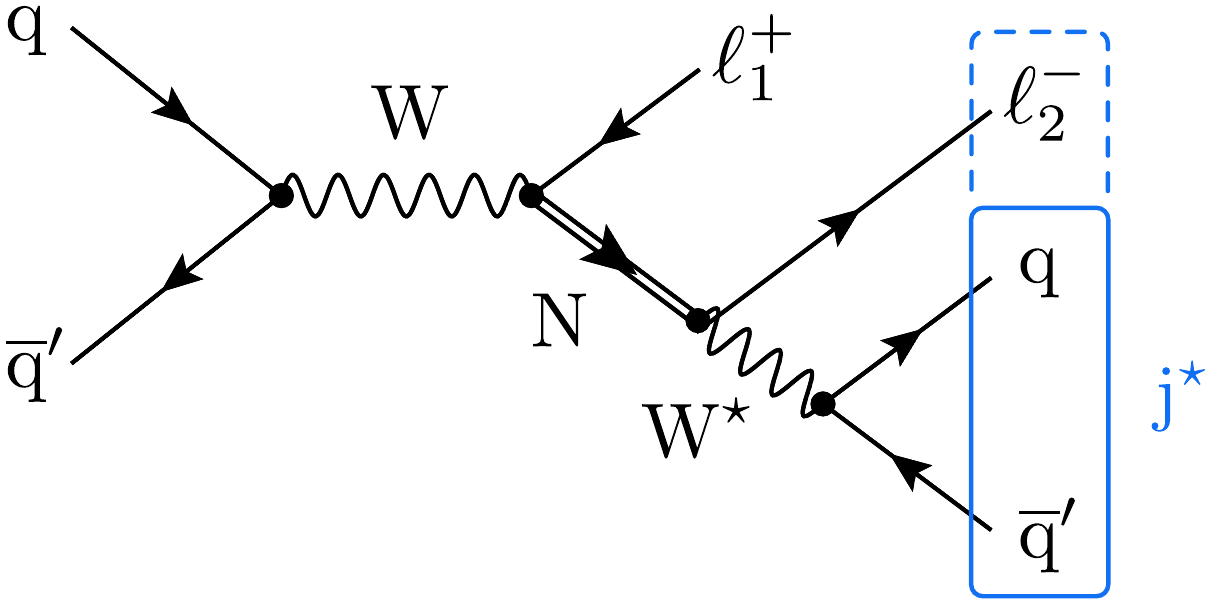}
\caption{\label{fig:intro-feynman} Born-level Feynman diagram for Dirac HNL production and decay via a charged-current interaction. Corresponding diagrams exist also for Dirac anti-HNL and Majorana HNL production and decay. In this search, the HNL decay products, encapsulated by the boxes, can be found collimated within a single jet \jn.}
\end{figure*}

The two most common three-body decay modes of HNLs involve a virtual \PW~boson and result in final states comprising two leptons and a neutrino or one lepton and two quarks with branching fractions of approximately 25 and 50\% for $\mhnl>5\GeV$, respectively. A signature considered in this search is depicted in Fig.~\ref{fig:intro-feynman}, which shows the Feynman diagram for Dirac HNL production at the LHC via an on-shell \PW~boson for $\mhnl<m_{\PW}$, and the subsequent HNL decay to a lepton and two quarks. Alternatively, HNLs can also be produced in decays of \PQb hadrons for $\mhnl<m_{\PQb}$ at the LHC, which is not considered in this search. For long-lived HNLs, the decay products are spatially separated (\ie displaced) from the point where the HNL is produced. Thus, the process depicted in Fig.~\ref{fig:intro-feynman} results in a signature comprising one prompt lepton \leptI, a displaced lepton \leptII, and at least one displaced jet \jn, a collimated set of particles originating from the hadronization of the two quarks. If the HNL is highly Lorentz-boosted, a single jet can also be formed encapsulating all displaced decay products from the HNL, a case that is also considered in this analysis. For Dirac HNLs, the final-state leptons have opposite electrical charges, while for Majorana HNLs they can have either the same or opposite electrical charges with equal probability.

In this paper, a search for long-lived Dirac and Majorana HNLs is performed without assuming that their coupling is exclusive to only one lepton generation. Instead, HNL production is investigated in the full coupling space spanned by \ve, \vmu, and \vtau for $2<\mhnl<20\GeV$, where an HNL proper lifetime $\tau_{0}$ of up to $\ctau=10^{4}\unit{mm}$ can be probed. A data set of proton-proton (\Pp{}\Pp{}) collisions recorded by the CMS detector at a centre-of-mass energy of 13\TeV and corresponding to an integrated luminosity of 138\fbinv is analyzed. Events containing two leptons (\Pe or \PGm) and jets are selected to target the hadronic HNL decay channel. Several event categories are defined to provide sensitivity to Dirac and Majorana HNL signals, both prompt and displaced, while contributions from background processes are estimated from data. A central feature of the analysis is the identification of displaced jets (with or without an overlapping displaced lepton) using a deep neural network (DNN) that does not explicitly require the reconstruction of displaced vertices. The DNN uses a domain adaptation technique to ensure accurate performance of the resulting classifier in data. The presented search explores an orthogonal phase space compared to Ref.~\cite{CMS:2022fut}, where similar parameter ranges in \mhnl and \ctau are probed but events with three leptons and no jets have been investigated instead.

An outline of the paper is as follows: a brief description of the CMS detector and the reconstruction of events is given in Section~\ref{sec:cms}. The real and simulated data sets are detailed in Section~\ref{sec:samples}, followed by a description of the event selection and categorization in Section~\ref{sec:selection}. In Section~\ref{sec:tagger}, details of the identification of displaced jets based on a DNN are presented including its architecture and training setup. The determination of the number of events from background processes is described in Section~\ref{sec:background}. A list of the sources of systematic uncertainties affecting the expected number of signal and background events can be found in Section~\ref{sec:systematics}. The results are presented in Section~\ref{sec:results}, followed by a summary in Section~\ref{sec:summary}. 

Tabulated results are provided in the HEPData record for this analysis~\cite{hepdata}.
 \section{The CMS detector and event reconstruction}
\label{sec:cms}

The central feature of the CMS apparatus is a superconducting solenoid of 6\unit{m} internal diameter, providing a magnetic field of 3.8\unit{T}. Within the solenoid volume are a silicon pixel and strip tracker, a lead tungstate crystal electromagnetic calorimeter (ECAL), and a brass and scintillator hadron calorimeter (HCAL), each composed of a barrel and two endcap sections. Forward calorimeters extend the pseudorapidity $\eta$ coverage provided by the barrel and endcap detectors. Muons are detected in gas-ionization chambers embedded in the steel flux-return yoke outside the solenoid. A more detailed description of the CMS detector, together with a definition of the coordinate system used and the relevant kinematic variables, can be found in Ref.~\cite{Chatrchyan:2008zzk}.

The particle-flow (PF) algorithm~\cite{Sirunyan:2017ulk} aims to reconstruct and identify each individual particle in an event, with an optimized combination of information from the various elements of the CMS detector. The energy of photons is obtained from the ECAL measurement. The energy of electrons is determined from a combination of the electron momentum at the primary interaction vertex as determined by the tracker, the energy of the corresponding ECAL cluster, and the energy sum of all bremsstrahlung photons spatially compatible with originating from the electron track. The energy of muons is obtained from the curvature of the corresponding track. The energy of charged hadrons is determined from a combination of their momentum measured in the tracker and the matching ECAL and HCAL energy deposits, corrected for the response function of the calorimeters to hadronic showers. Finally, the energy of neutral hadrons is obtained from the corresponding corrected ECAL and HCAL energies. The PF algorithm is able to reconstruct electrons and muons with efficiencies of about 80 and 95\%, respectively, for transverse displacements of less than 1\unit{cm}. At larger displacements, the efficiencies gradually reduce and approach zero at 40\unit{cm} and 1\unit{m}, respectively, as determined from simulation.

The missing transverse momentum vector \ptvecmiss is defined as the projection onto the plane perpendicular to the beams of the negative vector momentum sum of all PF candidates in an event. Its magnitude is referred to as \ptmiss. Jets are reconstructed from PF candidates and clustered by applying the anti-\kt algorithm~\cite{Cacciari:2008gp} with a distance parameter of 0.4. The reconstruction of jets is fully efficient with transverse displacements up to the edge of the inner tracking volume. After this, the reconstruction efficiency degrades as displaced charged particles can only be identified through their calorimetric energy deposits.

The primary vertex (PV) is taken to be the vertex corresponding to the hardest scattering in the event, evaluated using tracking information alone, as described in Section 9.4.1 of Ref.~\cite{CMS-TDR-15-02}.

Events with spuriously high \ptmiss, originating from a variety of reconstruction failures, detector malfunctions, or noncollision backgrounds, are vetoed by event filters with negligible impact on events with genuine \ptmiss~\cite{Sirunyan:2019kia}.

Events of interest are selected using a two-tiered trigger system. The first level, composed of custom hardware processors, uses information from the calorimeters and muon detectors to select events at a rate of around 100\unit{kHz} within a fixed latency of 4\mus~\cite{CMS:2020cmk}. The second level, known as the high-level trigger, consists of a farm of processors running a version of the full event reconstruction software optimized for fast processing, and reduces the event rate to around 1\unit{kHz} before data storage~\cite{CMS:2016ngn}.

 \section{Data set and simulated samples}
\label{sec:samples}

The analyzed sample of $\Pp\Pp$ collision data was recorded at $\sqrt{s}=13\TeV$ with the CMS detector during 2016, 2017, and 2018. The subsamples by year correspond to integrated luminosities of 36.3, 41.5, and 59.7\fbinv, respectively. The 2016, 2017, and 2018 data sets are analyzed independently and only combined for the final result to account for differences related to the data-taking conditions and to apply proper corrections and calibrations per year. Events were triggered by requiring at least one electron or muon candidate within $\abs{\eta}<2.4$. Electron candidates must have a transverse momentum \pt of at least 27 (32)\GeV in 2016 (2017, 2018), and pass additional identification requirements~\cite{Sirunyan:2020ycc}. Muon candidates have to fulfil $\pt>24\GeV$ (2016, 2018) or $\pt>27\GeV$ (2017), and have to be isolated from other activity in the event~\cite{CMS:2021yvr}.

Samples of simulated signal events are used to evaluate the detector resolution, efficiency, and acceptance for HNL production processes. Additionally, samples of simulated background events are used for understanding the composition of the data and to optimize the analysis strategy. Samples of simulated events are furthermore used to train machine learning algorithms. However, the final results are obtained by determining the contribution from background processes using data in sideband regions.

Samples of HNL production with various values of \mhnl ranging from 2 to 20\GeV, and with $10^{-5}<\ctau <10^{4}\unit{mm}$, are generated in the five-flavour scheme at leading order (LO) accuracy (\ie tree-level) using the \MGvATNLO v2.6.5 event generator~\cite{Alwall:2014hca}. The parton distribution function (PDF) set used is NNPDFv3.1~\cite{NNPDF:2017mvq}. Parton showering is performed with \PYTHIA{}~8.230~\cite{Sjostrand:2014zea} using the underlying event tunes CUETP8M1~\cite{Khachatryan:2015pea} (2016) or CP5~\cite{Sirunyan:2019dfx} (2017, 2018). Each sample is generated assuming an equal coupling strength for the three lepton generations, $\ve=\vmu=\vtau$. The resulting events are reweighted to also probe different combinations of coupling strengths. For these, the HNL displacement, determined by the total HNL decay width $\Gamma_{\hnl}$, has to be kept constant such that the subsequent detector simulation and event reconstruction do not need to be rerun. The HNL decay width can be written as
\begin{equation}
\Gamma_{\hnl}(\mhnl,\ve,\vmu,\vtau) =\underbrace{A_\Pe (\mhnl) \abs{\ve}^{2}}_{=\,\Gamma_{\Pe}} + \underbrace{A_{\PGm} (\mhnl) \abs{\vmu}^{2}}_{=\,\Gamma_{\PGm}} + \underbrace{A_{\PGt} (\mhnl) \abs{\vtau}^{2}}_{=\,\Gamma_{\PGt}}, \label{eq:samples-width}
\end{equation}
where the factors $A_{\lept}$, calculated by \MGvATNLO, describe the dependence of the partial widths per lepton flavour $\Gamma_{\lept}$ on \mhnl. Equation~\eqref{eq:samples-width} describes an ellipsoid spanned by \ve, \vmu, and \vtau for a given $\Gamma_{\hnl}$ and \mhnl. Thus, the generated events of a particular sample can be reweighted to any coupling combination on the ellipsoid surface. The barycentric coordinates \fe, \fmu, and \ftau with $\sum \fl=1$ are introduced, which denote the relative ratios
between the coupling strengths squared $\abs{\ve}^{2}:\abs{\vmu}^{2}:\abs{\vtau}^{2}\equiv \fe:\fmu:\ftau$. In total, 66 equidistant weights in \fl space are used, which have been calculated by the event generator at the matrix element level~\cite{Mattelaer:2016gcx}. The modelling of low-\pt jets originating from HNL decays has been validated by using the average charged particle multiplicity and comparing it to previous measurements in \Pep{}\Pem{} annihilation~\cite{Ali:1979em}. The HNL production cross sections are calculated at next-to-leading order (NLO) in the strong coupling constant, using the same generator settings, and are found to be approximately 10\% higher than the cross sections calculated at tree level for all samples.

Further details on the HNL Lagrangian and the implementation of models for both Dirac and Majorana HNLs can be found in Refs.~\cite{Atre:2009rg,Alva:2014gxa,Degrande:2016aje,Pascoli:2018heg}.

The major background processes are the production of vector bosons in association with jets (\wjets, \zjets, and \vgjets, where $\PV\in\{\PW,\PZ\}$), top quark production (\ttbar, single \PQt), and multijet events consisting of jets produced through the strong interaction. Samples of simulated background events are generated at NLO accuracy using the \MGvATNLO or \POWHEG~\cite{Frixione:2007nw,Alioli:2010xd,Frederix:2012dh} 2.0 event generators with the exception of multijet events that are generated at LO with \PYTHIA. The NNPDFv3.1 set is used and parton showering is performed by \PYTHIA with the same underlying event tunes as used for the signal samples.

The simulated events are overlaid with additional simulated minimum bias collisions (``pile\-up'') according to the distributions inferred from the data per year. All generated events undergo a full \GEANTfour~\cite{geant} simulation of the detector response and are subjected to the same event reconstruction as the data.
 \section{Event selection and categorization}
\label{sec:selection}

Events containing two leptons ($\lept\in\{\Pe,\PGm\}$) and at least one jet are analyzed. This signature targets the production and decay of HNLs via charged currents, $\Pp\Pp\to\leptI\hnl$, $\hnl\to\leptII\PQq\PAQq'$. The first lepton \leptI is expected to originate from the production vertex, while the second lepton \leptII and the jet(s) from the hadronization of the two quarks can be displaced with respect to the PV. The hadronic HNL decay channel targeted by this analysis has a branching fraction of about 50\% for $\mhnl >5\GeV$, which is approximately twice as large compared to the leptonic decay channel, $\hnl\to\leptpm\leptmp\PGn$. Other decay channels occur via neutral currents and result in $\PGn\PQq\PAQq$ or $\PGn\PGn\PAGn$ final states. Decays to other BSM particles, \eg lighter HNLs, are not considered in the model used for generating the signal samples. In scenarios where $\mhnl\lesssim 8\GeV$, the HNL is found to be sufficiently Lorentz-boosted with respect to the laboratory frame such that its decay products are collimated inside a single jet. The selected events also include leptonic \PGt decays to \Pe or \PGm. Tau lepton decays to hadrons are not considered in this analysis because efficient event trigger strategies with sufficiently low \pt thresholds on \PGt candidates were not available. In the following, the selection criteria for the physics objects are described.

\subsection{Physics object selection}

Two types of candidates are defined for electrons and muons depending on whether they are considered for \leptI or \leptII. The former is used to trigger the event and thus has to fulfil more stringent, tight identification criteria, while the latter is required to pass only relaxed, loose identification criteria instead. This increases the signal selection efficiency in cases when \leptII is displaced and/or clustered inside a jet.

Tight electron candidates are required to have a \pt of at least 29\,(34)\GeV and be within $\abs{\eta}<2.4$ for the 2016 (2017, 2018) data-taking periods. A boosted decision tree (BDT) discriminant is used to identify genuine electrons with an efficiency of ${\approx}90\%$ as measured in Drell--Yan events~\cite{Sirunyan:2020ycc}. This BDT is trained using several features of the electron track and of the associated energy deposits in the ECAL, including the relative isolation parameter $\eiso$. This parameter is defined as the scalar \pt sum of the charged and neutral PF candidates within a cone of radius $\Delta R = \sqrt{\smash[b]{(\Delta\eta)^2+(\Delta\phi)^2}}=0.3$ around the electron candidate, excluding the candidate itself, divided by the electron \pt. Here, $\Delta\eta$ and $\Delta\phi$ are, respectively, the pseudorapidity and azimuthal angle measured in radians and relative to the electron direction. The relative contribution from pileup is estimated as $A_\text{eff}\,\rho$ and subtracted from the isolation parameter, where $A_\text{eff}$ denotes an $\eta$-dependent effective area, and $\rho$ is the median of the transverse energy density in a $\delta\eta$-$\delta\phi$ region calculated using the charged particle tracks associated with the pileup vertices. Furthermore, the electron candidate has to originate from the vicinity of the PV by requiring a transverse and longitudinal impact parameter of $\abs{\dxy}<0.05\,(0.10)\unit{cm}$ and $\abs{d_\mathrm{z}}<0.1\,(0.2)\unit{cm}$, respectively, in the ECAL barrel (endcap) region.

Loose electron candidates need to fulfil $\pt>5\GeV$, $\abs{\eta}<2.4$, and pass a series of loose selection criteria that have been optimized to enhance the acceptance for displaced electrons that can also be clustered within nearby jets, as defined below. The criteria are a modified version of the standard identification requirements~\cite{Sirunyan:2020ycc} but allow for any number of missing electron track hits in the inner tracker. Furthermore, no requirements on the compatibility with the PV or on $\eiso$ are imposed.

Tight muon candidates must be within $\abs{\eta}<2.4$, have a \pt of at least 26, 29, 26\GeV for the 2016, 2017, 2018 data-taking periods, respectively, and fulfil additional identification requirements optimized for the selection of genuine muons that result in efficiencies of ${>}96\%$ as determined from a data sample of Drell--Yan events~\cite{Sirunyan:2018fpa}. In particular, muon candidates are required to be within $\abs{\dxy}<0.01\unit{cm}$ and $\abs{d_\mathrm{z}}<0.05\unit{cm}$ of the PV. Furthermore, the candidates must be isolated with a relative isolation parameter $\muiso<15\%$, which is defined as the scalar \pt sum of PF photons and charged and neutral PF hadrons within a cone of radius of $\Delta R = 0.4$, divided by the muon \pt. Contributions from pileup are estimated from the energy deposited by charged hadrons within the isolation cone that are associated with pileup vertices and subtracted from the isolation parameter.

Loose muon candidates are required to pass $\pt>3\GeV$,  $\abs{\eta}<2.4$, and fulfil only minimal requirements necessary for being reconstructed by the PF algorithm~\cite{Sirunyan:2018fpa}. Similar to loose electron candidates, loose muon candidates are allowed to be inside jets and do not have to be compatible with the PV.

Events with exactly two leptons of any charge and flavour combination are considered for further analysis, where at least one needs to fulfil the tight candidate requirements. This results in event categories, denoted ``\leptI{}\leptII'', where the first lepton is the tight candidate that is also matched to a corresponding event trigger and has to have $\pt(\leptI)>\pt(\leptII)$ to avoid the double counting of events where both leptons are responsible for a positive decision by the trigger system.

The momentum of each jet is defined as the vectorial momentum sum over all its constituent PF candidates. To mitigate the contribution of additional tracks and calorimetric energy depositions from pileup interactions, PF candidates identified as originating from pileup vertices are discarded and an offset correction is applied to account for remaining contributions~\cite{Khachatryan:2016kdb}. Jet energy corrections are derived from simulation so that their average measured energy becomes identical to that of particle level jets. In situ measurements of the momentum balance in dijet, photon+jet, Z+jet, and multijet events are used to determine any residual differences between the jet energy scale in data and in simulation, and appropriate corrections are made. The corrections are propagated to the measured \ptvecmiss. 

The analysis considers jets within $\abs{\eta}<2.4$ with a calibrated \pt greater than 20\GeV. Jets that overlap with \leptI within $\Delta R<0.4$ are ignored, whereas the jet \pt needs to be at least 30\GeV if \leptII is within $\Delta R<0.4$ from the jet. Jets close to \leptII are also ignored if $I_{\text{rel}}(\leptII)<15\%$ to reject jets that are primarily clustered from isolated leptons. The jet that is the closest in $\Delta R$ to \leptII is interpreted as the jet from the (displaced) HNL decay and is hereafter referred to as \jn. In simulation this procedure assigns a reconstructed jet correctly to the HNL decay with an efficiency of larger than 95\%. Events with more than one displaced jet from the HNL decay passing the event selection constitute a negligible fraction and are hence not explicitly targeted by the selection.

\subsection{Event categories}

The invariant mass \mllj is calculated from the summed four momenta of the two leptons and \jn. In signal events, \mllj approximates the mass of the \PW boson. The signal region (SR) is defined for selected events with an invariant dilepton mass of $20<\mll<80\GeV$, $\ptmiss<60\GeV$, and consisting of 1--4 jets, to reject \zjets, \wjets, and high-\pt processes (\eg{} \ttbar) with many jets that can additionally be produced in association with neutrinos. Additionally, \jn must be within $\Delta R<1.3$ of \leptII since in signal events both are expected to be spatially close together and even overlap when the HNL is sufficiently boosted. To suppress contributions from low-\pt leptons produced during hadronization, which are often found in \wjets and multijet events, the ratio $\pt(\jn)/\pt(\leptII)$ must be smaller than 2--4 depending on the data-taking period and category, since the \pt threshold for leptons also varies per data-taking period. Additionally, $\PGmpm{}\PGmmp{}$ events must have $\mll<70\GeV$ to further reject the overwhelming \zjets background in these categories. After applying all object and event selection criteria, the following approximate fractions of signal events remain when summed over all lepton flavour combinations: 2.5\% for $\ctau\leq 1\unit{mm}$, 1.5\% for $\ctau=10\unit{mm}$, 0.9\% for $\ctau=100\unit{mm}$, and less than 0.1\% for scenarios with larger \ctau values. These low efficiencies are primarily due to HNLs decaying into low-\pt leptons that often do not satisfy the \pt thresholds applied as part of the event selection. 

A control region (CR), enriched in background events (mainly \zjets), is obtained by inverting the SR selection on $\mll$, \ie to be larger than $80\GeV$. This region is used for validating various aspects of the analysis strategy such as the modelling of the simulation and the displaced jet tagger (Section~\ref{sec:tagger}).

Events in the SR are categorized through a series of criteria to be simultaneously sensitive to various HNL mass and coupling scenarios. First, events are separated depending on the lepton flavour combination (\Pe{}\Pe{}, \Pe{}\PGm, \PGm{}\Pe, \PGm{}\PGm) for probing different coupling scenarios and on the lepton charge combination, \ie opposite- or same-sign (OS, SS) events, to be sensitive to Dirac and Majorana HNL production, simultaneously. Furthermore, events are classified as either ``boosted'' if \leptII overlaps with \jn (\ie $\Delta R<0.4$) or ``resolved'' otherwise. Finally, events are split into three subcategories depending on the significance of the transverse displacement \dxysig of \leptII with respect to the PV; \dxysig is defined as $\dxy/\dxyerr$, where \dxy and \dxyerr are the transverse impact parameter and its uncertainty, respectively, as determined by the lepton track fit. The $\dxy/\dxyerr$ parameter is chosen because potential mismeasurements in \dxy and \dxyerr cancel each other out in the ratio and its distribution in the CR is found to be well described by the simulation. Events are grouped by the $\dxysig$ values into prompt-like ($\dxysig < 3$), displaced ($3 < \dxysig < 10$), and very displaced ($\dxysig > 10$) subcategories that probe a wide range of HNL lifetime scenarios. In total, this results in 48 independent event categories.

\begin{figure*}[htb!]
\centering
\includegraphics[width=0.48\textwidth]{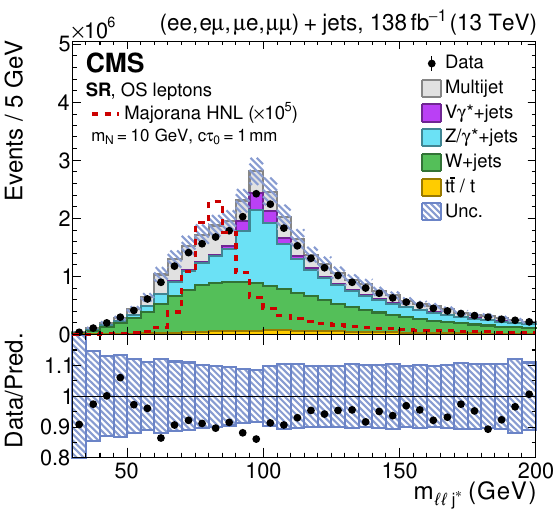}\hspace{0.03\textwidth}
\includegraphics[width=0.48\textwidth]{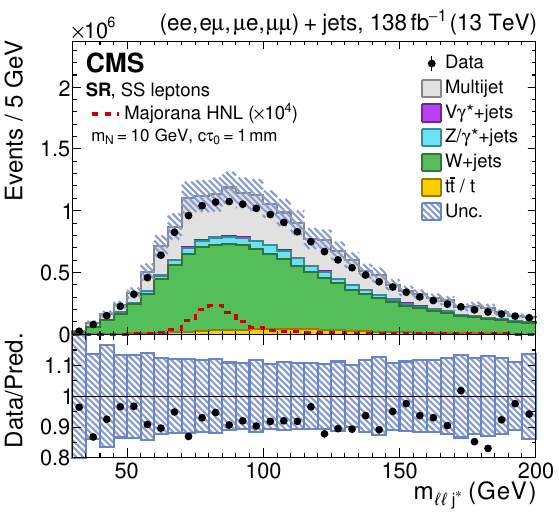}
\caption{\label{fig:selection-mllj}Distributions of \mllj for events with (left)~OS leptons and (right)~SS leptons in the signal region. A representative signal scenario for Majorana HNL production with equal coupling to all lepton generations is overlaid with its expected cross section scaled up as indicated in parentheses. The hatched band shows the total experimental systematic uncertainty in the simulated background prediction including the uncertainty from the finite sample size. The bottom panel shows the ratio of data over the prediction.}
\end{figure*}

Distributions of \mllj in the SR are presented in Fig.~\ref{fig:selection-mllj} for events with OS and SS lepton pairs. The data are found to be described well by the prediction from simulation normalized to the integrated luminosity. A representative signal scenario for Majorana HNL production assuming $\mhnl=10\GeV$ and $\ctau=1\unit{mm}$ with $\ve=\vmu=\vtau$ is overlaid. A resonance is expected in the \mllj distribution for signal events at the \PW boson mass, while a broader spectrum is found for background processes. The background in events with OS lepton pairs consists primarily of \zjets and \wjets events, while \vgjets, \ttbar, single top quark, and multijet processes are subdominant. A peak around the \PZ boson mass is seen in \zjets events since \mllj is mainly calculated from the two leptons momenta when the hadronic activity in \jn is low. In SS lepton pair events, multijet and \wjets events get selected because of misidentified leptons and compose the majority of the background. The number of multijet events appears to be slightly overestimated by the simulation resulting in an overall mismatch in normalization between data and prediction, which is however found to be covered by uncertainties. For further analysis, only events in the signal-enhanced region, defined by $70<\mllj<90\GeV$, are kept, whereas the sideband region in \mllj is used to estimate the background contributions from data (Section~\ref{sec:background}). Overall, the total background is of comparable size in both categories and since the Majorana and Dirac HNL production cross sections are identical, a similar sensitivity for both HNL types is expected.

 \section{Identification of displaced jets}
\label{sec:tagger}

Displaced jets that originate from the HNL decay products are identified using a jet tagger based on a DNN. The tagger is trained using jets originating from HNL decays with a transverse displacement \Lxy between 100\unit{$\mu\mathrm{m}$} and 1\unit{m}, where the lower threshold corresponds to a factor 2--4 times the position resolution of the PV~\cite{CMS:2014pgm}. The DNN is trained on various features of the selected jets and parameterized with respect to \Lxy, resulting in an optimized multiclass classifier that can be applied for a large variety of jets and displacements. The network is based on the developments reported in Ref.~\cite{Sirunyan:2019nfw}, but its setup has been extended in several aspects to account for the specific properties of jets from HNL decays, as detailed in the following. 

The DNN architecture consists of approximately 330\,000 parameters. The set of input features includes global jet features, as well as specific features for each type of jet constituent. The global jet features are the jet \pt, $\abs{\eta}$, mass, and area; the fraction of energy carried by the constituents within cones of $\Delta R=0.1,0.2,0.3,0.4$ with respect to the jet axis; the number of constituents carrying 60 or 90\% of the jet energy; the charged, electromagnetic, muon, and electron energy fractions; the variables used in the combined secondary vertex b tagging algorithm~\cite{Sirunyan:2017ezt}; the 1-, 2-, and 3-subjettiness variables~\cite{Thaler:2010tr}; and various event shape observables that are calculated from the constituents in the rest frame of the jet to probe its substructure. In total, 47 global jet features are used.

The features per constituent are grouped into blocks according to their type as follows:

\begin{itemize}
\item \textit{Charged PF candidates}: These candidates have an associated track and can thus provide information about the potential displacement of the jet within the tracking volume. The features encapsulate various properties of the track, its displacement with respect to the PV, and whether the track is also linked to a secondary vertex or a PF electron or muon. Up to 25 charged PF candidates are considered per jet, which are sorted according to the significance of their displacement in the transverse plane with respect to the PV. In total, 36 features are used per charged PF candidate.

\item \textit{Neutral PF candidates}: Features from up to 25 neutral constituents are used as input, which are important for identifying jets from HNL decays that occur at the outer edge of the tracking volume, where the track reconstruction efficiency is low. The neutral candidate features include the likelihood that the candidate stems from pileup interactions as calculated by the PUPPI algorithm~\cite{Bertolini:2014bba}. The candidates are sorted by either their distance in $\Delta R$ to a secondary vertex or by their \pt in cases where no secondary vertex has been reconstructed. In total, 9 features are used per neutral PF candidate.

\item \textit{Secondary vertices}: Features from up to four secondary vertices, reconstructed with the inclusive secondary vertex finder algorithm~\cite{Khachatryan:2011wq}, that are found within $\Delta R<0.4$ of the jet axis and use at least one jet constituent in their reconstruction, are provided as input to the DNN as well. They are sorted by the significance of their displacement in the transverse plane with respect to the PV. In total, 14 features are used per secondary vertex.

\item \textit{PF electrons/muons}: If a charged PF candidate is linked to a PF electron or muon, features that are specific to their reconstruction are used as additional inputs. For electrons, the properties of the Gaussian-sum-filter track~\cite{Adam_2005} and matching clusters in the ECAL are used. For muons, the features include details of the global track fit and the hit pattern in the muon chambers. Up to two electrons and muons are considered per jet, which are sorted according to the significance of their displacement in the transverse plane with respect to the PV. In total, 78 (37) features are used per electron (muon) candidate.
\end{itemize}

In the first stage of the DNN, the various jet constituent features are compressed through a series of one-dimensional convolutions with varying filter sizes using a kernel size of one. The last convolutional layer can be seen as a bottleneck that forces the DNN to extract only the most discriminating information from the larger number of input features, which is then passed on to the next stage. The compressed features of each block are flattened and concatenated with global jet features and \Lxy, where the latter acts as a parameter of the DNN~\cite{Baldi:2016fzo}. The result is then passed through two dense layers with 200 nodes. At this stage, the network splits into one branch to predict the jet class and into another branch to predict the jet domain (\ie simulation or data), where each branch consists of another series of dense layers. The following jet classes are considered: jets from the hadronization of gluons or quarks; jets from pileup interactions; jets clustered primarily from prompt \Pe, \PGm, photons, or hadronically decaying \PGt leptons; displaced jets from the HNL decay with or without displaced \Pe, \PGm, or hadronically decaying \PGt leptons.

In general, the intricate modelling of displaced signatures is less well studied compared to prompt ones. Hence, features of displaced jets can possess a different trend in data compared to simulation. The distribution of output discrimination scores from a classifier, trained solely on simulated events, can differ significantly when evaluated on simulation or data. In a previous study, differences of up to 50\% have been seen between the selection efficiencies of data and simulation~\cite{Sirunyan:2019nfw}. In this search, a domain adaptation technique is used to mitigate such differences and effectively calibrate the response of the tagger for simulated events. Technically, a layer is inserted into the domain branch that reverses the gradients of the loss function with respect to the DNN parameters in the preceding layers. During the training, this forces the DNN to retain only discriminating features from the inputs that are domain-invariant~\cite{ganin2015unsupervised}, \ie features that are well described by the simulation. The loss function is chosen to approximate the earth mover distance~\cite{arjovsky2017wasserstein}, which yields well-defined gradients and thus stabilizes the adversarial training. The relative strength of the domain loss is chosen such that a trade-off is achieved between obtaining a discriminant with approximately equal efficiencies in data and simulation while simultaneously retaining high discrimination between the various jet classes.

The class branch is trained using jets from independent samples of simulated events from various processes. The jets are resampled to obtain a balanced set with equal class proportions as a function of the jet \pt and $\eta$. Random \Lxy values are assigned to background jets that are sampled from the distribution of \Lxy values of true HNL jets~\cite{Baldi:2016fzo}. The domain branch is trained on jets from data and simulated background events in the CR, where the latter are weighted according to the expected cross section per process.

\begin{figure*}[hp!]
\centering
\includegraphics[width=0.48\textwidth]{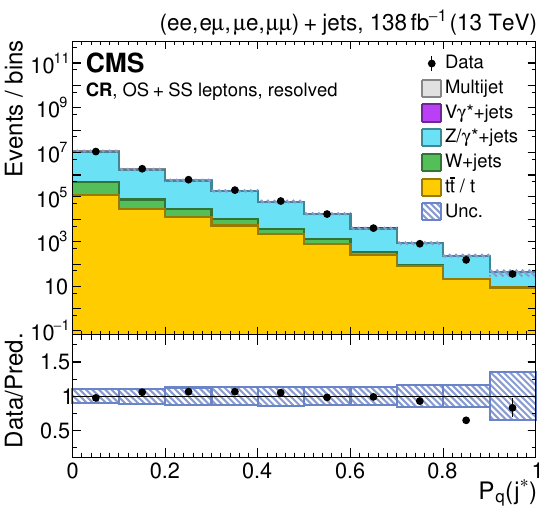}\hspace{0.03\textwidth}
\includegraphics[width=0.48\textwidth]{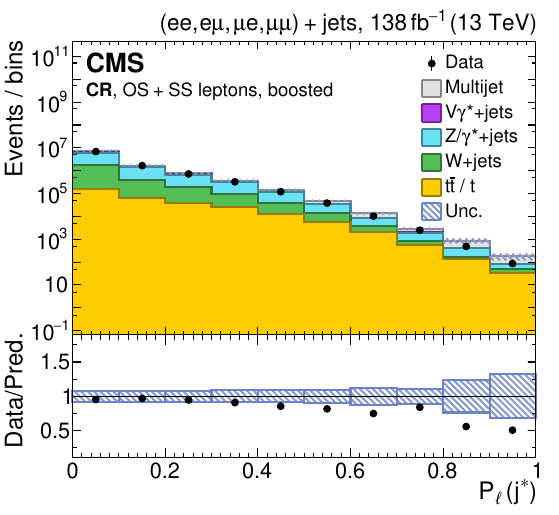}\\
\includegraphics[width=0.48\textwidth]{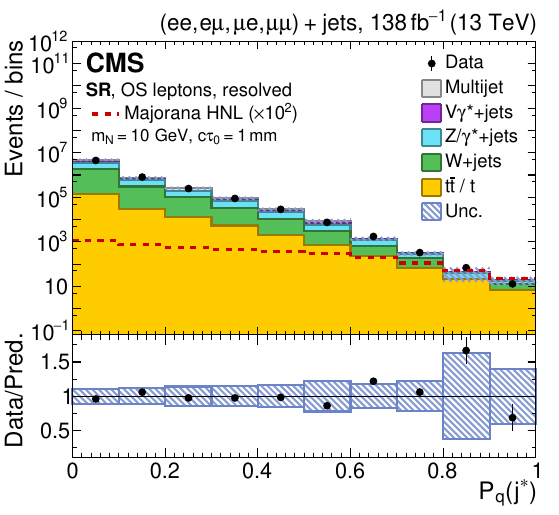}\hspace{0.03\textwidth}
\includegraphics[width=0.48\textwidth]{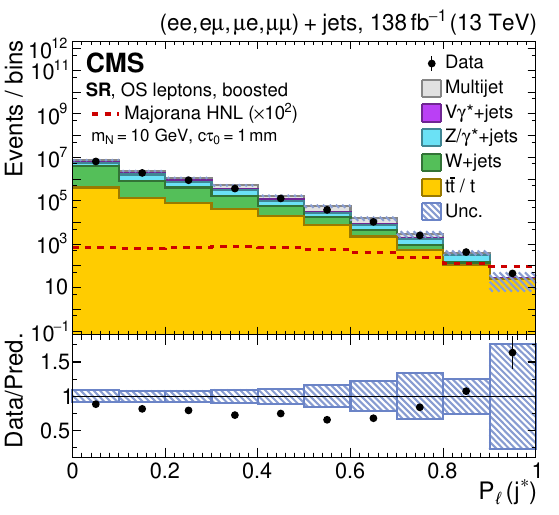}\\
\includegraphics[width=0.48\textwidth]{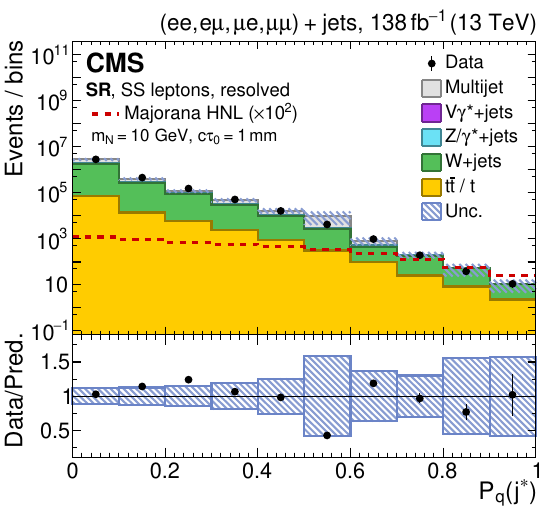}\hspace{0.03\textwidth}
\includegraphics[width=0.48\textwidth]{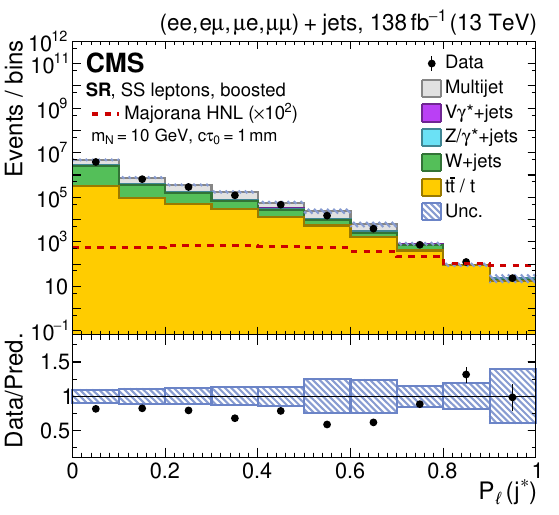}
\caption{\label{fig:tagger-crsr}Distributions of the displaced jet tagging score for (left)~resolved and (right)~boosted categories: (upper row)~control region; (middle row)~signal region with OS leptons; (lower row)~signal region with SS leptons. A representative signal scenario for Majorana HNL production with equal coupling to all lepton generations is overlaid with its expected cross section scaled up as indicated in parentheses. The hatched band shows the total experimental systematic uncertainty in the simulated background prediction including the uncertainty from the finite sample size. The bottom panel shows the ratio of data over the prediction.}
\end{figure*}

The best discrimination against SM jets is found by calculating the ratios for each displaced jet likelihood over the sum of the SM background likelihoods. The ratios are monotonically mapped to the range 0--1 for simplicity. Per displaced jet, the \Lxy parameter is chosen in the range $10^{-1}$--$10^{3}$\unit{mm} to maximize the resulting score for each class. For true displaced jets, the score will be the largest when \Lxy matches the displacement of the HNL, whereas it is randomly distributed for SM jets. The resulting scores are used as follows. For resolved events, the score for displaced jets with and without hadronically decaying \PGt leptons, \pq, is evaluated. On the other hand, for boosted events the flavour of \leptII determines whether the score for displaced jets with electrons, \pe, or with muons, \pmu, is evaluated instead. Distributions of the resulting scores are shown in Fig.~\ref{fig:tagger-crsr} for resolved and boosted categories in the CR and SR. In general, the prediction from simulation, normalized to the integrated luminosity, describes the data well in all regions, except for a slight overestimation of multijet events as seen before in Fig.~\ref{fig:selection-mllj}. Selections on the resulting displaced-jet tagger score can reduce the SM backgrounds by several orders in magnitude. For example, a selection of $\pj>0.6$ reduces the background yield by a factor of about 3000\,(800) while retaining 10\,(17\%) of signal events for boosted (resolved) events considering the signal scenario in Fig.~\ref{fig:tagger-crsr}. In the analysis, optimized thresholds on \pql are applied ranging from 0.2--0.7 depending on the event category to achieve the best sensitivity, as described later in Section~\ref{sec:results}.

 \section{Estimation of background processes}
\label{sec:background}

Various background sources contribute to the SR, with different relative fractions depending on the category. A substantial background contribution originates from \zjets production in events with two genuine OS leptons of same flavour. In events with either SS leptons or OS leptons of different flavour, the dominant source of background events is misidentified leptons, which primarily originate from \wjets, \vgjets, and multijet production, whereas events from \ttbar production are less prominent. For final states containing electrons, ``charge-flip'' backgrounds, where the charge of one of the electrons is misidentified, are also possible.

Contributions from background processes are determined from data to accurately account for the reconstruction effects, selection efficiencies, and object calibrations of background events in the final event categories. This procedure has the additional advantage that no large samples of background events have to be simulated and subsequently calibrated for modelling the background contribution in the SR.

\begin{figure*}[htb!]
\centering
\includegraphics[width=0.48\textwidth]{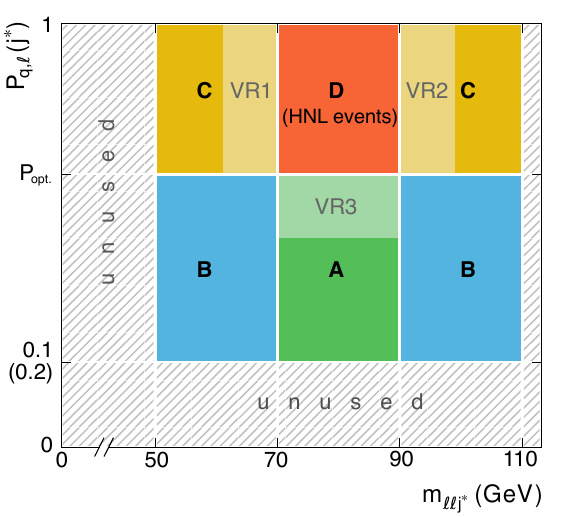}
\caption{\label{fig:background-vr}Schematic of the orthogonal regions in the (\mllj,\pql) plane, that are used to determine the background in region D from data through an ABCD method. The threshold $P_\text{opt.}$ depends on the category. Regions with $\pql<0.1\,(0.2)$ for resolved (boosted) events are not considered. The three VRs that are subsets of regions A and C are indicated in lighter shades. The figure is not to scale.}
\end{figure*}

The yields in data, $n_{X}^{\alpha}$, from three background-enriched sideband regions $\alpha\in\{\mathrm{A,B,C}\}$ in each of the 48 categories $X$ per data-taking period are used to determine the background contribution, $b_{X}^\mathrm{D}$, in a signal-enriched region. The regions are defined by selecting events to be either above or below optimized thresholds on the \pql score or on the \mllj observable, which are found to be independent from studies of simulated events. These selections partition the SR into four orthogonal regions as sketched in Fig.~\ref{fig:background-vr}. 

A so-called matrix  (``ABCD'') method is used, where the pass-fail ratio of an event passing the threshold on \pql is measured in \mllj sidebands (50--70\GeV or 90--110\GeV) and then applied to the observed yield found within 70--90\GeV. The yields $b_{X}^\mathrm{D}$ are determined through a maximum likelihood fit to data using the following model to properly account for non-Gaussian uncertainties in case of small event yields:
\begin{multline}
\label{eq:bkg-ll}
    L\left(r, \vec{b}, \vec{\theta}, \vec{\kappa} \middle|\, \vec{n}\right) =  \prod_{i}^\text{2016,2017,2018}~\prod_{j}^\text{resolved,boosted}~\prod_{k}^\text{subcategory}   
    \prod_{\alpha}^\mathrm{A,B,C}
    \\
    \operatorname{Poisson}\left(n_{ijk}^{\alpha} \middle|\, b_{ijk}^{\alpha} + r \, s_{ijk}^{\alpha} (\vec{\theta} )\right)  
     \operatorname{Poisson}\left(n_{ijk}^\mathrm{D} \middle|\, b_{ijk}^\mathrm{D}  + r \, s_{ijk}^\mathrm{D} (\vec{\theta} )\right) p(\vec{\theta}),\quad  b_{ijk}^\mathrm{D}\equiv\kappa_{ij} \frac{b_{ijk}^\mathrm{B} \, b_{ijk}^\mathrm{C}}{b_{ijk}^\mathrm{A}},
\end{multline}
where $n_{ijk}^{\alpha}$ is the number of events observed in region $\alpha$ with $i$ denoting the data-taking period, $j$ denoting the event topology, and $k$ denoting the subcategory $\{\Pe\Pe,\Pe\PGm,\PGm\Pe,\PGm\PGm\}\otimes\{\text{OS,SS}\}\otimes\{\dxysig(\leptII)~\text{bins}\}$.  Nuisance parameters $\vec{\theta}$ that control the effects of systematic uncertainties are constrained through prior probability density functions $p(\vec{\theta})$. A possible signal contamination of $r \,s_{ijk}^{\alpha}$ in the sidebands is explicitly considered, where $r$ modifies the expected signal yields $s_{ijk}^{\alpha}$ from simulation for the given HNL scenario under consideration. 

In addition to the signal strength $r$, fit parameters $\kappa_{ij}$ are introduced, for which the fitted values will be found to diverge from unity in the presence of a bias from the ABCD method. Separate $\kappa_{ij}$ parameters are used per data-taking period and per event topology. In summary, the parameters of this fit are $r$, $b_{ijk}^{\alpha}$, $\vec{\theta}$, and $\kappa_{ij}$.

The validity of the background estimation method is tested by predicting the data yields in three background-dominated validation regions (VRs) that are defined as follows. The first two VRs comprise events within \mllj ranges of 65--70\GeV or 90--100\GeV together with selections on the \pql scores. For the third VR, events within a \mllj range of 70--90\GeV that are within an interval in the \pql scores below $P_\text{opt.}$ are used. The three VRs in the (\mllj,\pql) plane are sketched in Fig.~\ref{fig:background-vr}. Another test is performed in an additional VR, defined using a narrow \mllj window within 5\GeV of region D, comprising subsets of events from both VR1 and VR2, to test the background estimation method close to region D.

Per VR, each $\kappa_{ij}$ parameter is determined through a simultaneous fit to 24 independent event categories. Overall, the determined event yields agree well with the observations in all VRs and uncertainties of 10--25\% are assigned on $b_{ijk}^\mathrm{D}$, chosen based on the most significant deviations of $\kappa_{ij}$ from unity. These uncertainties are implemented in fits of the SR through log-normal priors on the $\kappa_{ij}$ parameters.
 \section{Systematic uncertainties}
\label{sec:systematics}

The expected numbers of signal and background events can be affected by several sources of systematic uncertainties. Their impact is modelled through nuisance parameters that change the expected events yields and are correlated across all categories and years, where applicable. Most uncertainties only impact the signal prediction since it is obtained from simulation, whereas the number of expected background events is determined from data in sideband regions, as detailed in Section~\ref{sec:background}.

A set of systematic uncertainties related to the selection efficiencies and calibration of leptons, jets, and \ptmiss is evaluated for simulated events as described below. These are propagated through the displaced jet tagger by recalculating the scores on the modified inputs where applicable.

Systematic uncertainties in the prompt lepton reconstruction, identification, and trigger efficiencies are accounted for by varying \pt and $\eta$-dependent normalization factors, derived from data-to-simulation ratios, within their uncertainties. Systematic uncertainties in the identification efficiency of the displaced lepton candidates are taken into account by varying additional scale factors within their uncertainties. These have been derived in $\PJGy\to\mu\mu$ events and in asymmetric $\PGg\to\Pe\Pe$ conversion events. Furthermore, an uncertainty in the lepton track reconstruction efficiency is considered, derived from tracks in \ttbar-enriched events.

The jet energy scale in simulated events is modified for all jets using the uncertainty associated with the \pt and $\eta$-dependent scale factors. The resulting differences are propagated to \ptmiss. The jet energy resolution is modified in simulated events by either increasing or decreasing the difference in \pt with respect to the matched jet at the generator level. Random smearing is used when no generator-level jet is found within $\Delta R<0.4$. The component of \ptmiss that is not clustered inside jets, and thus remains uncorrected by the jet energy scale, is varied within its uncertainty.

An uncertainty of 5\% in the total inelastic cross section~\cite{CMS:2018mlc} is used when calculating the expected distribution of the number of pileup interactions in data.

An uncertainty in the displaced track reconstruction and jet tagging efficiency, which accounts for differences between data and simulation, is determined using tracks matched to secondary vertices in events from a \ttbar-enriched region. These differences are parameterized as a function of the track $\abs{\dxy}$. A correction and the corresponding uncertainty in the displaced jet tagging efficiency is derived by varying the efficiency of the three most significantly displaced tracks within \jn. These are found to be less than 10\% for jets with prompt-like tracks ($\dxy<0.1\unit{cm}$) and in the range 10--20\% at larger displacements. Uncertainties in the mistagging efficiency for jets from SM backgrounds are accounted for through the uncertainties in the background estimation.

The factorization and renormalization scales are varied by factors of 2 and 0.5 in simulated signal samples using precomputed weights by the event generator~\cite{Kalogeropoulos:2018cke}. Only residual changes after the selection are taken as uncertainties while the overall effect on the normalization is considered part of the theoretical uncertainty instead. Uncertainties due to the PDF choice are determined by reweighting each event according to the prescribed variations of the corresponding PDF set.

The integrated luminosities of the 2016, 2017, and 2018 data-taking periods are individually known with uncertainties in the 1.2--2.5\% range~\cite{CMS-LUM-17-003,CMS-PAS-LUM-17-004,CMS-PAS-LUM-18-002}, while the total 2016--2018 integrated luminosity has an uncertainty of 1.6\%, the improvement in precision reflecting the (uncorrelated) time evolution of some systematic effects. 

Uncertainties resulting from the finite size of the simulated samples are incorporated into the likelihood model using the Barlow--Beeston method~\cite{Barlow:1993dm}.

The background yields are determined in-situ using the likelihood model defined in Eq.~\eqref{eq:bkg-ll}, which properly accounts for the statistical uncertainties in the data yields in the sidebands. Additional uncertainties in the range of 10--25\% in the background yield are considered to account for the difference in the background estimation per data-taking period and for additional differences between resolved and boosted categories.

In summary, the statistical uncertainties from the sideband regions in the background yields are found to have the largest impact on the determined signal yield for the signal+background hypothesis, which can be as large as 10\% depending on the category. Smaller impacts originate from the loose muon reconstruction (7\%), the jet tagging efficiency (6\%), the loose electron reconstruction (4\%), and the jet energy scale and resolution (2\%).
 \section{Results}
\label{sec:results}

\begin{figure*}[htb!]
\centering
\includegraphics[width=0.48\textwidth]{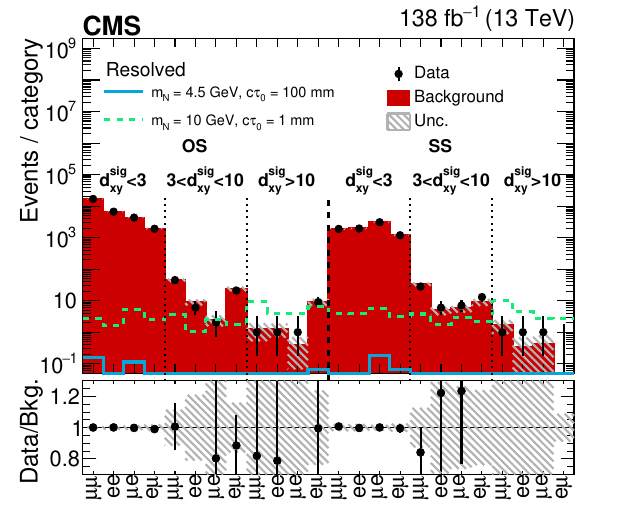}\hspace{0.03\textwidth}
\includegraphics[width=0.48\textwidth]{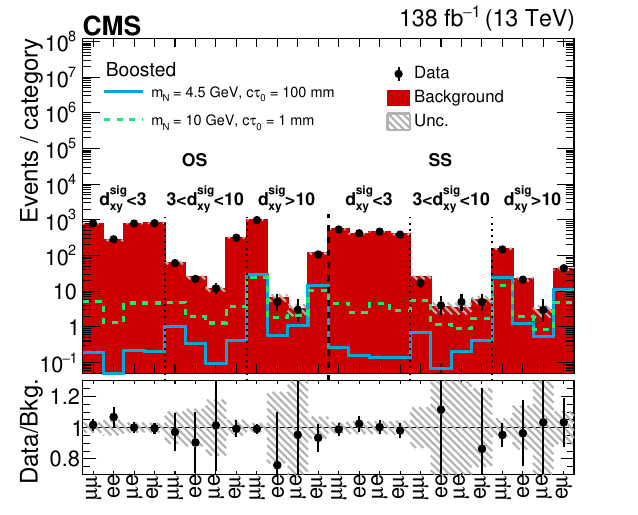}
\caption{\label{fig:results-observed-data}Observed number of events and predicted number of background events per category for (left)~resolved and (right)~boosted categories. The bin label denotes the flavour of the prompt (\leptI) and displaced (\leptII) lepton as \leptI{}\leptII. Two representative signal scenarios for Majorana HNL production with equal coupling to all lepton generations are overlaid. The lower panels show the ratio of the data to the predicted background. The hatched band shows the total systematic uncertainty in the predicted background.}
\end{figure*}

The thresholds on the \pql scores are optimized per category to yield the best discovery sensitivity~\cite{gdiscov} to the various HNL scenarios as follows. The expected yields and uncertainties from background processes are determined from data in sideband regions and the expected signal yields are taken from simulation, such that no bias with respect to the observed data can occur. The optimal thresholds are found to be within the range 0.2--0.7, depending on the event category. The resulting background yields are compared to the observed number of events in Fig.~\ref{fig:results-observed-data}. The events in data are found to agree well with the estimated number of background events in each category.

Upper limits on the HNL production cross sections at 95\% confidence level (\CL) are determined as functions of \mhnl and the absolute values of the couplings strengths squared $\abs{\ve}^2$, $\abs{\vmu}^2$, and $\abs{\vtau}^2$. A modified frequentist approach is employed, which uses the profile likelihood ratio as the test statistic~\cite{CMS-NOTE-2011-005}, the \CLs criterion~\cite{junk, read}, and the asymptotic formulae~\cite{Cowan:2010js} to approximate the distributions of the test statistic under the background-only and signal-plus-background hypotheses.

\begin{figure*}[hp!]
\centering
\includegraphics[width=0.48\textwidth]{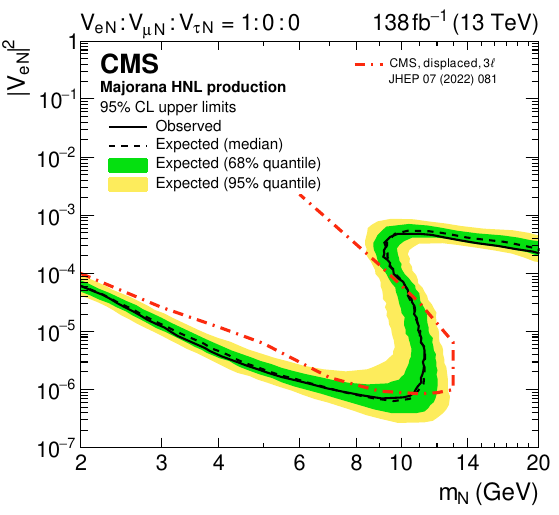}
\includegraphics[width=0.48\textwidth]{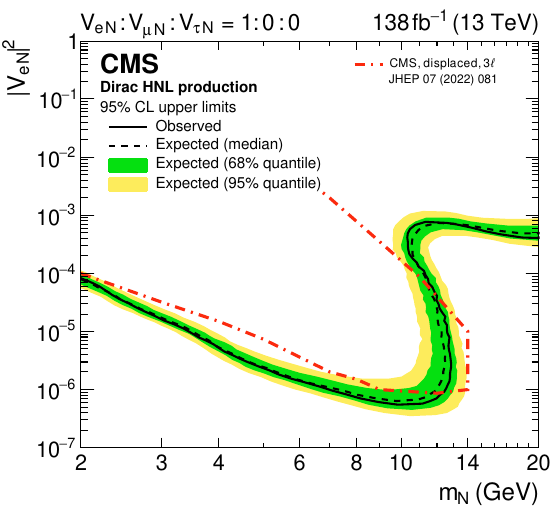}\\
\includegraphics[width=0.48\textwidth]{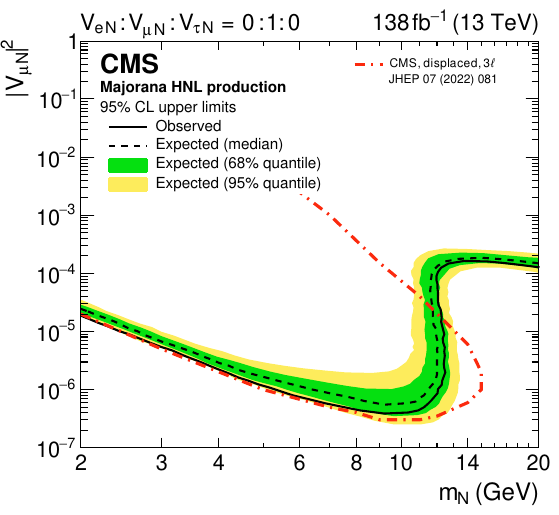}
\includegraphics[width=0.48\textwidth]{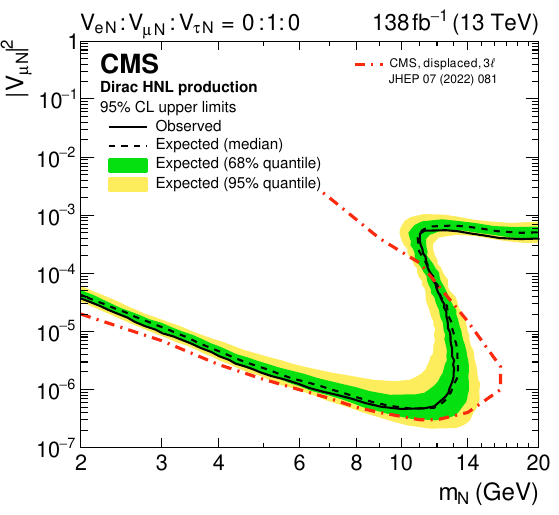}\\
\includegraphics[width=0.48\textwidth]{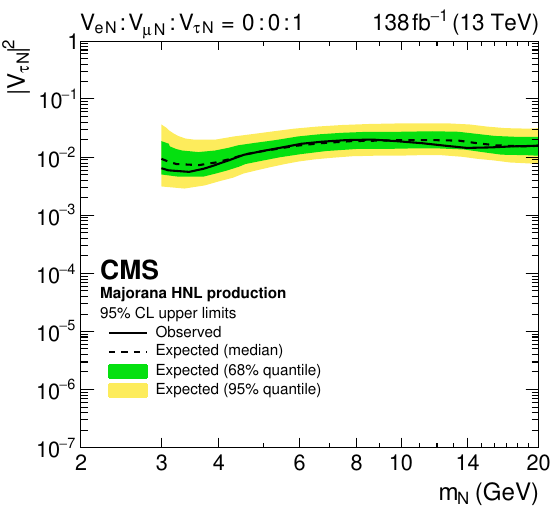}
\includegraphics[width=0.48\textwidth]{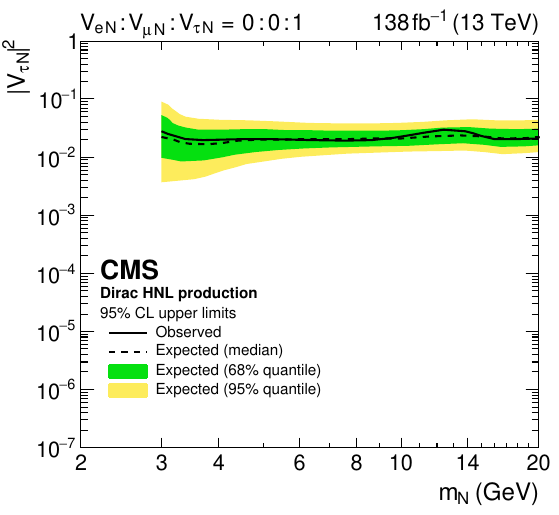}
\caption{\label{fig:results-pure-coupling}Expected and observed 95\% \CL limits on (left) Majorana and (right) Dirac HNL production as functions of the HNL mass and coupling strengths for (upper row) pure electron, (middle row) pure muon, and (lower row) pure \PGt lepton coupling scenarios. For the last coupling case no limits are derived for $\mhnl\lesssim3\GeV$ because of the tau mass. The dashed-dotted line shows the result of an orthogonal CMS analysis targeting HNL decays to leptons.}
\end{figure*}

\begin{figure*}[hp!]
\centering
\includegraphics[width=0.48\textwidth]{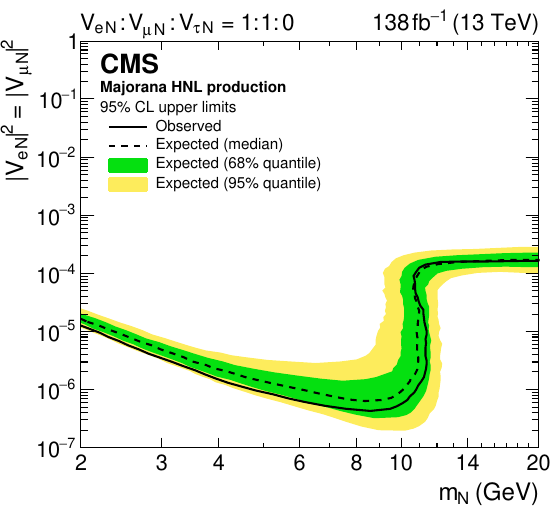}
\includegraphics[width=0.48\textwidth]{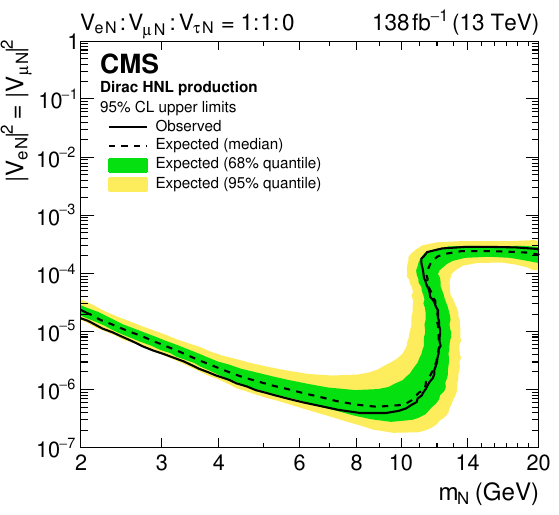}\\
\includegraphics[width=0.48\textwidth]{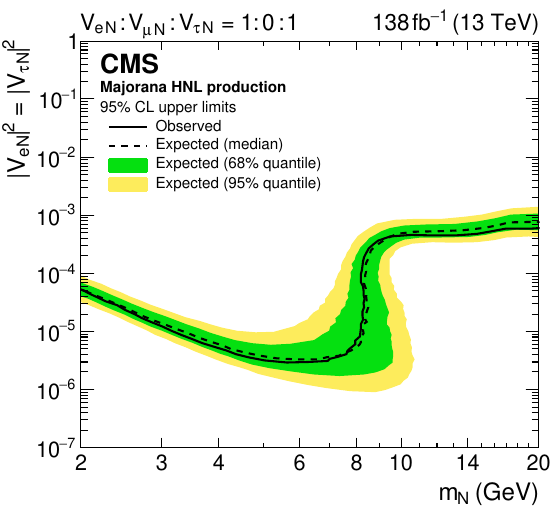}
\includegraphics[width=0.48\textwidth]{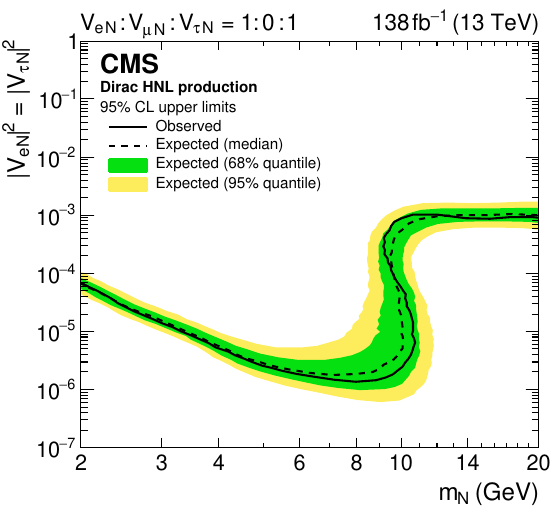}\\
\includegraphics[width=0.48\textwidth]{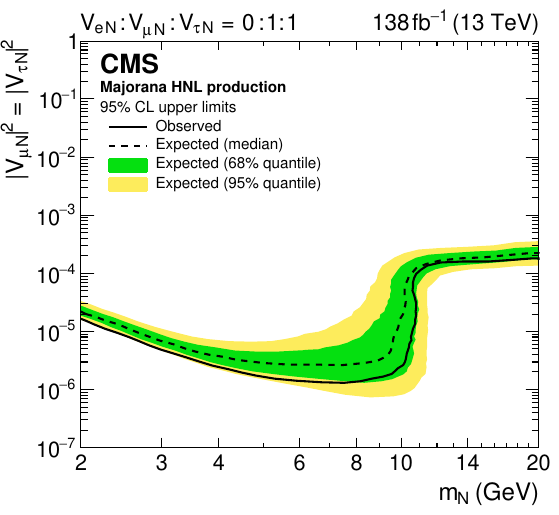}
\includegraphics[width=0.48\textwidth]{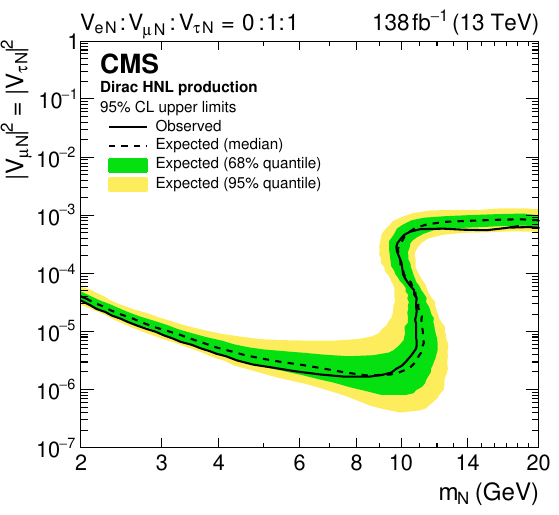}\\
\caption{\label{fig:results-mixed-coupling}Expected and observed 95\% \CL limits on (left) Majorana and (right) Dirac HNL production for various coupling scenarios as functions of the HNL mass and coupling strengths: (upper row)~mixed \Pe-\PGm couplings; (middle row)~mixed \Pe-\PGt couplings; (lower row)~mixed \PGm-\PGt couplings. The relative ratios of the coupling per lepton generation are indicated in the plots.}
\end{figure*}

\begin{figure*}[htb!]
\centering
\includegraphics[width=0.48\textwidth]{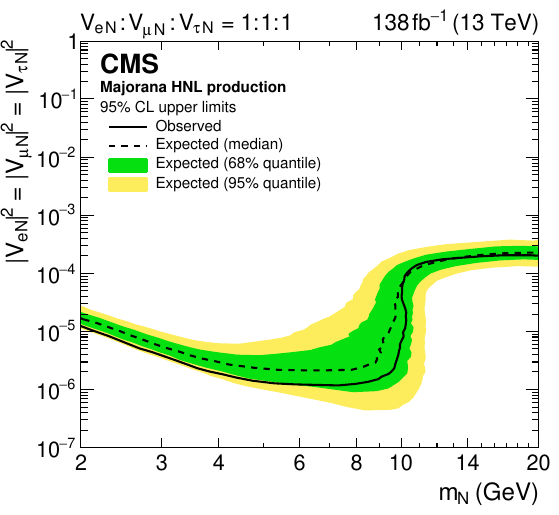}
\includegraphics[width=0.48\textwidth]{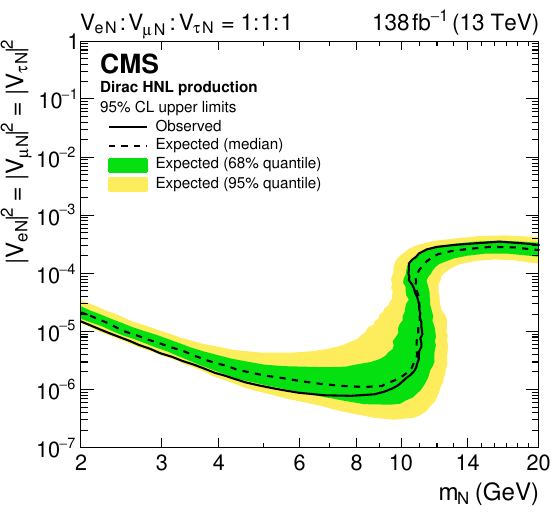}
\caption{\label{fig:results-democratic-coupling}Expected and observed 95\% \CL limits on (left) Majorana and (right) Dirac HNL production for the democratic couplings to all lepton generation scenario as functions of the HNL mass and coupling strengths.}
\end{figure*}

Two-dimensional exclusion limits on Dirac and Majorana HNL production are shown in  Figs.~\ref{fig:results-pure-coupling}, \ref{fig:results-mixed-coupling}, and \ref{fig:results-democratic-coupling} for scenarios of pure coupling, of mixed coupling, and of democratic couplings to all lepton generations, respectively. A broad parameter space in mass and coupling values is excluded by this search. For $\mhnl\lesssim 8\GeV$, most exclusion limits roughly follow a constant \ctau value. At higher masses, the cross section becomes too low and, at the same time, sensitivity to displaced scenarios is lost causing the limit curve to turn. At $\mhnl\gtrsim 14\GeV$, the limits have transitioned into a prompt regime and follow a constant cross section value with negligible dependence on \mhnl for $m_{\ell,\PQq}\ll\mhnl\ll m_{\PW}$. The sensitivity rapidly deteriorates towards the pure \PGt lepton coupling scenario since $\PGt\PGt$+jets events are only selected in the analysis in the case of leptonic \PGt decays to \Pe or \PGm, whereas \PGt lepton decays to hadrons are not considered. No limits can be derived for scenarios with $\mhnl\lesssim 3\GeV$ that are close to the pure \PGt lepton coupling case, because of the tau mass. For some HNL scenarios, the results are compared to an orthogonal search exploiting $\hnl\to\leptpm\leptmp\PGn$ decays, where the two leptons form a displaced vertex~\cite{CMS:2022fut}, which prevents sensitivity to prompt scenarios, resulting in a different trend in the limits at high masses. In this search, the sensitivity to prompt scenarios is achieved by the displaced jet tagger, in part due to the absence of any requirement on the presence (or otherwise) of reconstructed secondary vertices. The best limits are obtained for pure muon coupling scenarios due to the excellent muon identification performance. Coupling strength values as low as $\abs{\vmu}^{2}>5\,(4)\times10^{-7}$ are excluded for Dirac (Majorana) HNLs with $\mhnl=10\GeV$ at 95\% \CL that correspond to proper decay lengths of 17\,(10)\unit{mm}.

\begin{figure*}[htb!]
\centering
\includegraphics[width=0.48\textwidth]{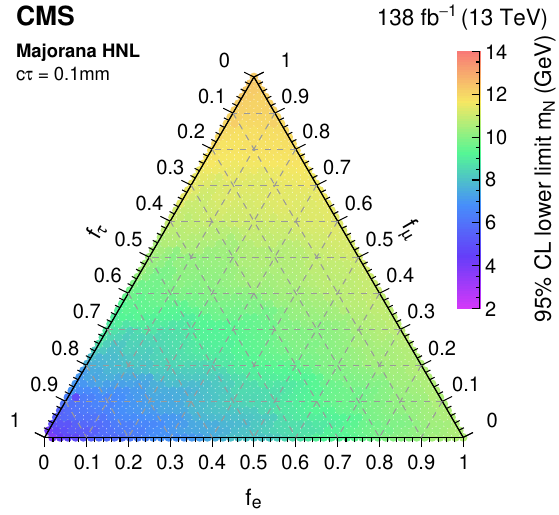}\hspace{0.03\textwidth}
\includegraphics[width=0.48\textwidth]{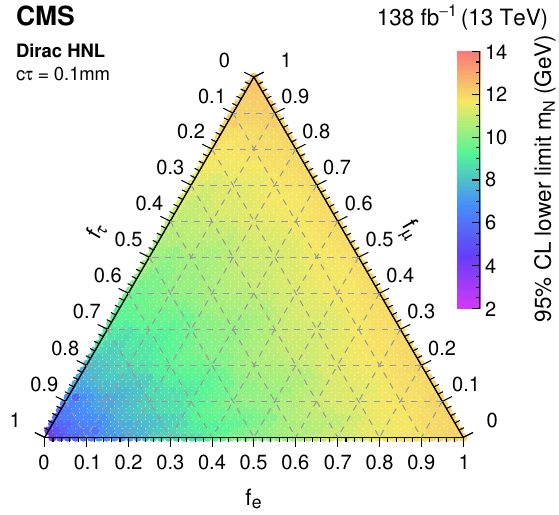}\\
\includegraphics[width=0.48\textwidth]{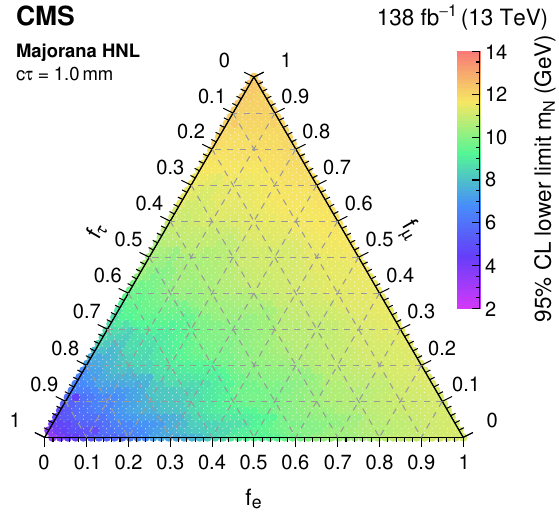}\hspace{0.03\textwidth}
\includegraphics[width=0.48\textwidth]{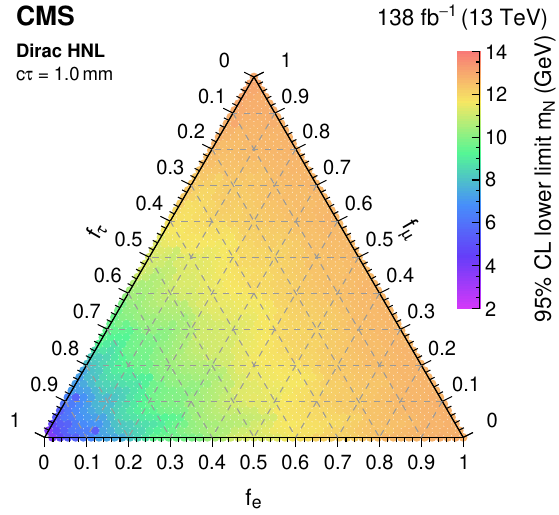}
\caption{\label{fig:results-triangle-mass}Observed 95\% \CL lower limits on the (left column) Majorana and (right column) Dirac HNL mass as functions of the relative couplings to the three lepton generations considering a fixed proper decay length of (upper row)~0.1\unit{mm} and (lower row)~1\unit{mm}. The limits are determined for $\mhnl>3\GeV$.}
\end{figure*}

\begin{figure*}[htb!]
\centering
\includegraphics[width=0.48\textwidth]{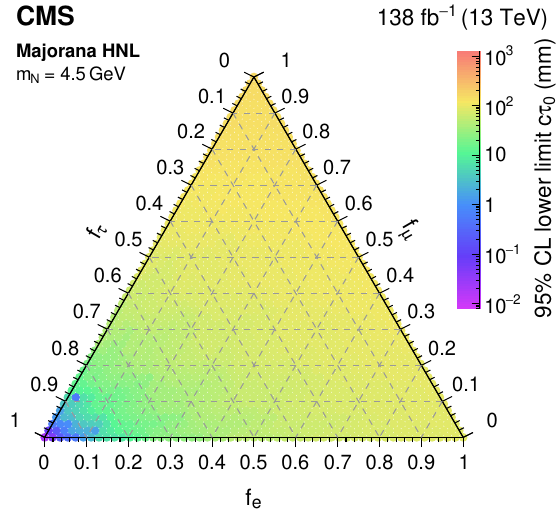}\hspace{0.03\textwidth}
\includegraphics[width=0.48\textwidth]{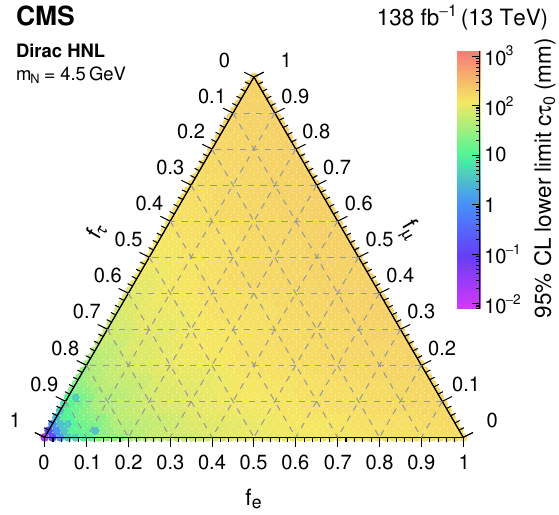}\\
\includegraphics[width=0.48\textwidth]{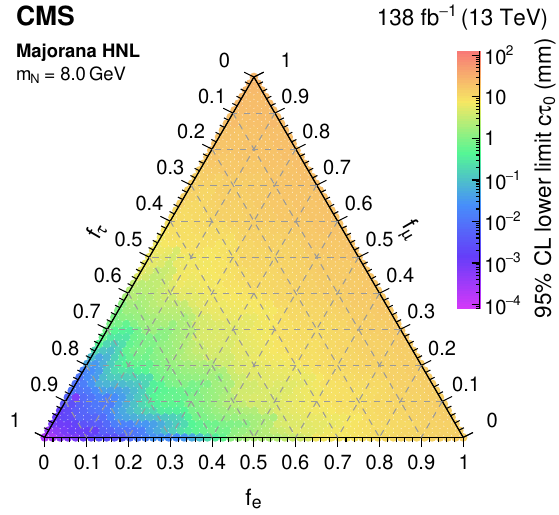}\hspace{0.03\textwidth}
\includegraphics[width=0.48\textwidth]{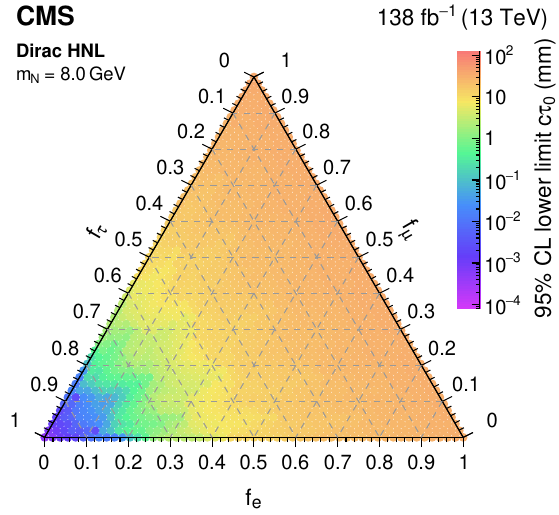}
\caption{\label{fig:results-triangle-ctau}Observed 95\% \CL lower limits on the (left column) Majorana and (right column) Dirac HNL proper decay length as functions of the relative couplings to the three lepton generations considering a fixed mass of (upper row)~4.5\GeV and (lower row)~8.0\GeV.}
\end{figure*}

Limits on the HNL mass and proper lifetime are presented in Figs.~\ref{fig:results-triangle-mass} and \ref{fig:results-triangle-ctau}, respectively, as functions of the relative coupling strengths to the three lepton generations, defined as $\fl=\abs{\vl}^{2}/(\abs{\ve}^{2}+\abs{\vmu}^{2}+\abs{\vtau}^{2})$. The results are obtained by interpolating the logarithmic value of the limits linearly between neighbouring points in $\fl$-space. The most stringent limits are found for pure muon coupling scenarios.

The results obtained from the hadronic HNL decay show a comparable sensitivity as previous HNL searches targeting the orthogonal leptonic final state by the ATLAS, CMS and LHCb Collaborations~\cite{ATLAS:2019kpx,ATLAS:2022atq,CMS:2018iaf,CMS:2018jxx,CMS:2022fut,LHCb:2020wxx,LHCb:2020akw}. In particular, the presented results for the pure electron coupling scenario are found to be more stringent for $\mhnl\lesssim4\GeV$. These are the first results reported that involve a prompt or long-lived Majorana or Dirac HNL that couples to all three lepton generations in the 2--20\GeV mass range.

 \section{Summary}
\label{sec:summary}

A search is presented for long-lived heavy neutral leptons (HNLs), which are predicted in extensions of the standard model of particle physics through a seesaw mechanism. Dirac or Majorana HNLs can mix with the three standard model lepton generations in a nontrivial way, resulting in lepton flavour number violation. Hence, a complete set of coupling scenarios involving all three lepton generations is considered. A novelty of this analysis is the simultaneous sensitivity to prompt and displaced scenarios through the use of a dedicated displaced jet tagger that does not explicitly rely on the presence of secondary vertices.

A data sample of proton-proton collision events recorded with the CMS experiment at the CERN LHC at a centre-of-mass energy of 13\TeV is analyzed, corresponding to an integrated luminosity of 138\fbinv. Events containing two leptons (electrons or muons) and jets are selected and categorized according to the lepton flavour and electrical charge, the displacement of the lower-momentum lepton, and whether the lepton overlaps with a nearby jet. Jets originating from the decay of a long-lived HNL are identified through a deep neural network using various features of the jet and its constituent particles. The network training relies on simulated event samples that cover a broad range of HNL masses and lifetimes, and acts as a powerful discriminant for jets that are spatially separated from the luminous region, even in the absence of reconstructed secondary vertices. 

Contributions from background processes are determined from data in sideband regions. No excess of events in the data over the expected background is observed. Upper limits on the HNL production cross section are determined as functions of the HNL mass and the three coupling strengths to each lepton generation. Exclusion limits are presented in the coupling-mass plane, as lower limits on the HNL lifetime, and on the HNL mass. Results are provided for Dirac and Majorana HNL production considering various coupling combinations. The most stringent limit on the coupling strength is obtained for pure muon coupling scenarios, with values of $\abs{\vmu}^{2}>5\,(4)\times10^{-7}$ excluded for Dirac (Majorana) HNLs with a mass of 10\GeV at a confidence level of 95\% that correspond to proper decay lengths of 17\,(10)\unit{mm}. 

This analysis is the first HNL search at the LHC that targets long-lived and hadronically decaying HNLs in the 2--20\GeV mass range, with inclusive coupling to all three lepton generations. The results show comparable sensitivity to an orthogonal analysis that targets long-lived HNLs decaying to leptons.

\begin{acknowledgments}
We congratulate our colleagues in the CERN accelerator departments for the excellent performance of the LHC and thank the technical and administrative staffs at CERN and at other CMS institutes for their contributions to the success of the CMS effort. In addition, we gratefully acknowledge the computing centres and personnel of the Worldwide LHC Computing Grid and other centres for delivering so effectively the computing infrastructure essential to our analyses. Finally, we acknowledge the enduring support for the construction and operation of the LHC, the CMS detector, and the supporting computing infrastructure provided by the following funding agencies: SC (Armenia), BMBWF and FWF (Austria); FNRS and FWO (Belgium); CNPq, CAPES, FAPERJ, FAPERGS, and FAPESP (Brazil); MES and BNSF (Bulgaria); CERN; CAS, MoST, and NSFC (China); MINCIENCIAS (Colombia); MSES and CSF (Croatia); RIF (Cyprus); SENESCYT (Ecuador); MoER, ERC PUT and ERDF (Estonia); Academy of Finland, MEC, and HIP (Finland); CEA and CNRS/IN2P3 (France); SRNSF (Georgia); BMBF, DFG, and HGF (Germany); GSRI (Greece); NKFIH (Hungary); DAE and DST (India); IPM (Iran); SFI (Ireland); INFN (Italy); MSIP and NRF (Republic of Korea); MES (Latvia); LAS (Lithuania); MOE and UM (Malaysia); BUAP, CINVESTAV, CONACYT, LNS, SEP, and UASLP-FAI (Mexico); MOS (Montenegro); MBIE (New Zealand); PAEC (Pakistan); MES and NSC (Poland); FCT (Portugal); MESTD (Serbia); MCIN/AEI and PCTI (Spain); MOSTR (Sri Lanka); Swiss Funding Agencies (Switzerland); MST (Taipei); MHESI and NSTDA (Thailand); TUBITAK and TENMAK (Turkey); NASU (Ukraine); STFC (United Kingdom); DOE and NSF (USA).

\hyphenation{Rachada-pisek} Individuals have received support from the Marie-Curie programme and the European Research Council and Horizon 2020 Grant, contract Nos.\ 675440, 724704, 752730, 758316, 765710, 824093, and COST Action CA16108 (European Union); the Leventis Foundation; the Alfred P.\ Sloan Foundation; the Alexander von Humboldt Foundation; the Science Committee, project no. 22rl-037 (Armenia); the Belgian Federal Science Policy Office; the Fonds pour la Formation \`a la Recherche dans l'Industrie et dans l'Agriculture (FRIA-Belgium); the Agentschap voor Innovatie door Wetenschap en Technologie (IWT-Belgium); the F.R.S.-FNRS and FWO (Belgium) under the ``Excellence of Science -- EOS" -- be.h project n.\ 30820817; the Beijing Municipal Science \& Technology Commission, No. Z191100007219010 and Fundamental Research Funds for the Central Universities (China); the Ministry of Education, Youth and Sports (MEYS) of the Czech Republic; the Shota Rustaveli National Science Foundation, grant FR-22-985 (Georgia); the Deutsche Forschungsgemeinschaft (DFG), under Germany's Excellence Strategy -- EXC 2121 ``Quantum Universe" -- 390833306, and under project number 400140256 - GRK2497; the Hellenic Foundation for Research and Innovation (HFRI), Project Number 2288 (Greece); the Hungarian Academy of Sciences, the New National Excellence Program - \'UNKP, the NKFIH research grants K 124845, K 124850, K 128713, K 128786, K 129058, K 131991, K 133046, K 138136, K 143460, K 143477, 2020-2.2.1-ED-2021-00181, and TKP2021-NKTA-64 (Hungary); the Council of Science and Industrial Research, India; ICSC -- National Research Centre for High Performance Computing, Big Data and Quantum Computing, funded by the EU NexGeneration program (Italy); the Latvian Council of Science; the Ministry of Education and Science, project no. 2022/WK/14, and the National Science Center, contracts Opus 2021/41/B/ST2/01369 and 2021/43/B/ST2/01552 (Poland); the Funda\c{c}\~ao para a Ci\^encia e a Tecnologia, grant CEECIND/01334/2018 (Portugal); the National Priorities Research Program by Qatar National Research Fund; MCIN/AEI/10.13039/501100011033, ERDF ``a way of making Europe", and the Programa Estatal de Fomento de la Investigaci{\'o}n Cient{\'i}fica y T{\'e}cnica de Excelencia Mar\'{\i}a de Maeztu, grant MDM-2017-0765 and Programa Severo Ochoa del Principado de Asturias (Spain); the Chulalongkorn Academic into Its 2nd Century Project Advancement Project, and the National Science, Research and Innovation Fund via the Program Management Unit for Human Resources \& Institutional Development, Research and Innovation, grant B37G660013 (Thailand); the Kavli Foundation; the Nvidia Corporation; the SuperMicro Corporation; the Welch Foundation, contract C-1845; and the Weston Havens Foundation (USA).
\end{acknowledgments}

\bibliography{auto_generated}

\cleardoublepage \appendix\section{The CMS Collaboration \label{app:collab}}\begin{sloppypar}\hyphenpenalty=5000\widowpenalty=500\clubpenalty=5000
\cmsinstitute{Yerevan Physics Institute, Yerevan, Armenia}
{\tolerance=6000
A.~Hayrapetyan, A.~Tumasyan\cmsAuthorMark{1}\cmsorcid{0009-0000-0684-6742}
\par}
\cmsinstitute{Institut f\"{u}r Hochenergiephysik, Vienna, Austria}
{\tolerance=6000
W.~Adam\cmsorcid{0000-0001-9099-4341}, J.W.~Andrejkovic, T.~Bergauer\cmsorcid{0000-0002-5786-0293}, S.~Chatterjee\cmsorcid{0000-0003-2660-0349}, K.~Damanakis\cmsorcid{0000-0001-5389-2872}, M.~Dragicevic\cmsorcid{0000-0003-1967-6783}, P.S.~Hussain\cmsorcid{0000-0002-4825-5278}, M.~Jeitler\cmsAuthorMark{2}\cmsorcid{0000-0002-5141-9560}, N.~Krammer\cmsorcid{0000-0002-0548-0985}, A.~Li\cmsorcid{0000-0002-4547-116X}, D.~Liko\cmsorcid{0000-0002-3380-473X}, I.~Mikulec\cmsorcid{0000-0003-0385-2746}, J.~Schieck\cmsAuthorMark{2}\cmsorcid{0000-0002-1058-8093}, R.~Sch\"{o}fbeck\cmsorcid{0000-0002-2332-8784}, D.~Schwarz\cmsorcid{0000-0002-3821-7331}, M.~Sonawane\cmsorcid{0000-0003-0510-7010}, S.~Templ\cmsorcid{0000-0003-3137-5692}, W.~Waltenberger\cmsorcid{0000-0002-6215-7228}, C.-E.~Wulz\cmsAuthorMark{2}\cmsorcid{0000-0001-9226-5812}
% ...
\par}
\vskip\cmsinstskip
\dag:~Deceased\\
$^{1}$Also at Yerevan State University, Yerevan, Armenia\\
$^{2}$Also at TU Wien, Vienna, Austria\\
% ...
\end{sloppypar}
\end{document}